\documentclass[aps,epsf,preprint,showpacs,superscriptaddress]{revtex4-1}
\usepackage{epsfig}

\usepackage{epsfig,pstricks,graphicx}
\usepackage{amsmath,amssymb}
\usepackage[mathscr]{eucal}
\usepackage{color}

\def\id{{\rm 1\kern-.22em l}}

\usepackage{amsmath}
\usepackage{amssymb}
\usepackage[T1]{fontenc}
\usepackage{graphicx}

\begin{document}

%
\title{Even-odd effect in the thermopower and strongly enhanced
       thermoelectric efficiency
       for superconducting single-electron transistors}

\author{Christopher Eltschka}
\affiliation{Institut f{\"u}r Theoretische Physik, Universit{\"a}t Regensburg,
             D-93040 Regensburg, Germany}
\author{Jens Siewert}
\affiliation{Departamento de Qu\'{\i}mica F\'{\i}sica, Universidad del Pa\'{\i}s Vasco -
             Euskal Herriko Unibertsitatea, 48080 Bilbao, Spain}
\affiliation{IKERBASQUE, Basque Foundation for Science, 48011 Bilbao, Spain}
  
\date{\today}

\begin{abstract}
It is well-known that the transport properties of 
single-electron transistors with a 
superconducting island and 
normal-conducting leads (NSN SET) 
may depend on whether or not there is a single
quasiparticle on the island. This parity effect
has pronounced consequences for the linear 
transport properties. Here we analyze the thermopower
of NSN SET with and without parity effect, for entirely
realistic values of device parameters. 
Besides a marked dependence of the thermopower on the
superconducting gap $\Delta$ we observe an enhancement 
in the parity regime
which is accompanied by a dramatic increase 
of the thermoelectric figure of merit $ZT$.
The latter can be explained within a simple re-interpretation
of $ZT$ in terms of averages and variances of transport
energies.

\end{abstract}


\maketitle


%

 \section{Introduction}

The properties of transport through small conducting islands 
have been investigated extensively during the past years. 
Electric current in such devices flows due to tunneling of single 
electrons and is subject to the so-called Coulomb blockade effect~\cite{Averin91}
which is characterized by a new energy scale, the capacitive
charging energy $E_C$ of the island (see below). 
In the recent past, substantial attention has been devoted also
to thermoelectric effects in single-electron devices~\cite{Beenakker92,Staring93,Dzurak97,Moeller01,Andreev01,Boese01,Matveev02,Turek02,Koch04,Turek2005,Kubala2008,Scheibner2008,Kuo2009,Costi2010,Triberis2010,Wierzbicki2011}.
Single-electron devices are interesting candidates for thermoelectric
applications as it has long been known that dimensional reduction of
the electron dynamics may lead to an enhanced thermoelectric 
efficiency~\cite{Dresselhaus1993,Mahan1996}. 

While an immense amount of work has been done to investigate thermopower
for quantum dots, surprisingly little is known about the thermoelectric effects
in single-electron transistors (SET) with superconducting electrodes.
In particular, SET
with superconducting islands are interesting as they may exhibit the
{\em parity effect} where a single unpaired quasiparticle  determines
the macroscopic thermodynamic properties of the island electrode~\cite{AverinNazarov1992,Tinkham1992} as well as the current-voltage characteristics of SET with normal-conducting
electrodes and a superconducting island (NSN SET)~\cite{Eiles1993,Hergenrother1994,GSZ1994}.
The parity effect can be observed below a crossover temperature $T^{\ast}\approx \Delta/8$
for typical system parameters (here $\Delta$ is the energy gap of the superconductor).
One may expect that the peculiar combination of properties like several competing
energy scales ($E_C$, $\Delta$, $T^{\ast}$) and the presence of a singularity in the 
quasiparticle spectrum gives rise to interesting behavior in the thermoelectric
response of such systems.

In Ref.~\cite{Turek2005} the thermopower for an NSN SET was studied for $\Delta < E_C$
and $T>T^{\ast}$. Even for this regime without parity effect, interesting oscillations
of the thermopower as a function of the electrostatic island potential 
and their strong dependence on the ratio $\Delta/E_C$ were predicted. In this article,
we investigate the thermopower of NSN SET for temperatures below the crossover
temperature. We find that the interplay of energy scales, Coulomb-blockade and parity
effects, and the peculiarity of the electronic spectrum lead to a rich variety 
of features in the thermopower $S$. Most intriguingly, however, for certain
gate voltages this system displays
a dramatic enhancement of the thermoelectric efficiency quantified by the figure of
merit $ZT$.

This paper is organized as follows. First we introduce the setup and the
theoretical methods that are used to describe transport in such systems.
Subsequently we briefly review the parity effect in SET with superconducting
islands. We then turn to describe the results for the thermopower, and to
interpret them in terms of average transport energies. Finally we discuss
the surprising results for the figure of merit $ZT$ which we explain
in the frame of a simple re-interpretation of this quantity.
%
%

\section{The transistor setup and master equation}
In an NSN SET, a superconducting island with a small electrostatic
capacitance $C$ is connected via tunnel junctions
to two normal-conducting leads (cf.~Fig.~1). The corresponding charging energy
$E_C\equiv e^2/(2C)\gg k_B T$ is large compared to the temperatures under
consideration (here, $e>0$ denotes the elementary charge). The conductances of
the tunnel junctions are assumed to be small compared to $e^2/h$, so
sequential tunneling dominates and cotunneling effects may be neglected.

\begin{figure}[h]
 \resizebox{.48\textwidth}{!}{\includegraphics{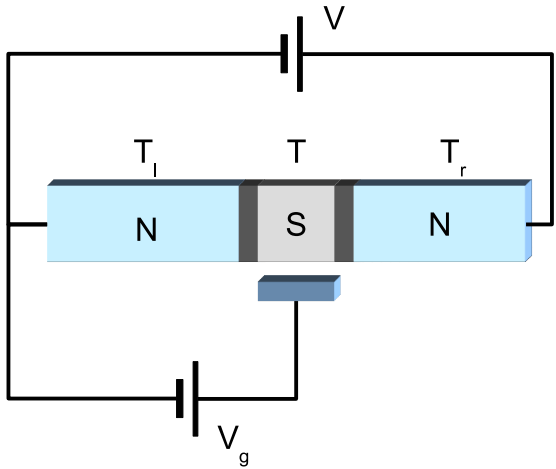}}
\caption{\label{fig:setup} 
  The NSN SET
  consists of a superconducting island (S) which
  is coupled to two
  normal-conducting leads (N) via tunnel barriers. The
  electrostatic potential of the island 
  can be controlled by the gate voltage $V_g$.
  There may flow a current through
  the system due to the bias 
  voltage $V$ or a temperature
  difference $\Delta T = T_l-T_r$ between the two leads.
  In order to detect the thermopower $S = -V/\Delta T$
  as a function of the gate voltage $V_g$
  the bias $V$ is adjusted such that the corresponding
  current exactly cancels the current which
  arises due to the temperature difference.
   }
\end{figure}
The electrostatic potential of the island can be controlled by means of an
external potential $n_x\propto V_g$ due to the gate voltage $V_g$, and 
the electrostatic energy of the setup with $n$ excess electrons on the island
can be expressed as
\begin{equation}
\label{eq:charging_energy}
E_n(n_x) = E_C \left( n^2 - 2 n  n_x \right) \ \ .
\end{equation}
The energy cost for adding a single electron to the island
is $u_n(n_x)=E_{n+1}(n_x)-E_n(n_x)$ while to add two electrons,
an energy $E_{n+2}(n_x)-E_n(n_x)=u_{n+1}(n_x)+u_{n}(n_x)$ is required.
A current may flow in the device if a bias voltage $\Delta V=V_r-V_l$
and/or a temperature difference $\Delta T=T_l-T_r$ is applied.
Throughout this work we consider the linear-response regime, that is,
$|\Delta V/E_C|\ll 1$ and $|\Delta T/T| \ll 1$.

Linear transport of charge and heat is conveniently described in terms
of the equations
\begin{equation}
\label{eq:lintransport}
   \left(\begin{array}{c}
           I_e \\
           I_q
         \end{array}
   \right)
=
   \left(\begin{array}{cc}
           G_V & G_T\\
           M   &  K
         \end{array}
   \right)
   \left(\begin{array}{c}
           \Delta V \\
           \Delta T
         \end{array}
   \right)
\end{equation}
which relate the charge and heat current response, $I_e$ and $I_q$, to
an applied potential and temperature difference. 
Here, $G_V$ is the linear (charge) conductance, and 
the thermal conductance $\kappa$ is defined via $I_q=\kappa\Delta T$
for $I_e=0$, {\em i.e.}, $\kappa=K-G_VTS^2$.
The relevant quantity for the thermoelectric response is
the thermopower $S$, given by $S=-\Delta V/\Delta T=G_T/G_V$.

In order to calculate the the conductances $G_V$, $G_T$ and the thermopower 
$S$ we employ a master-equation formalism.
According to the orthodox theory~\cite{Averin91} 
charge and heat current through the system can be written as
\begin{eqnarray}
\label{eq:ecurrent}
I_e &=&  - e \sum_n\sum_{j=1,2} 
                P_n \left[ \Gamma_r^{\, n\to n-j} - \Gamma_r^{\, n\to n+j} 
                   \right]
\\
\label{eq:qcurrent}
I_q &=&   \sum_n\sum_{j=1,2} 
                P_n \left[ q_r^{\, n\to n-j} - 
                            q_r^{\, n\to n+j} 
                   \right]
\end{eqnarray}
where $P_n$ is the stationary probability for finding $n$ excess electrons
on the island, $\Gamma_r^{n \to n-j}$ is the tunneling rate of $j$ electrons
from the island to the right lead, and $\Gamma_r^{n\to n+j}$ denotes the
tunneling rate of $j$ electrons from the right lead to the island.
Correspondingly, $q_r^{\, n\to n-j}$ is the energy transfer rate
in $j$-electron tunneling to the right lead whereas 
$q_r^{\, n\to n-j}$ denotes the energy transfer rate in a $j$-electron
tunneling event from the right lead to the island.
We consider only sequential tunneling and neglect co-tunneling events.

In Eqs.~\eqref{eq:ecurrent},\eqref{eq:qcurrent} we have taken into 
account the possibility of single-electron tunneling ($j=1$) and
coherent two-electron tunneling ($j=2$)~\cite{Hekking1993}.
The rates for the latter process are given by
\begin{equation}
\label{eq:2e-rate}
\Gamma_i^{\, n\to n \pm 2}(\epsilon^{(\pm 2)}_i)\ = \ \frac{G_{A,i}}{e^2}
                              \frac{\epsilon^{(\pm 2)}_i}{\exp{(\epsilon^{(\pm 2)}_i/k_B T_i)-1}}\ \ ,\ 
                               i=l\ , r 
\end{equation}
where $G_{A,l}$, $G_{A,r}$ are the Andreev conductances in the left and the
right junction and $\epsilon_l^{(\pm 2)}$, $\epsilon_r^{(\pm 2)}$ are the energies which are
dissipated in a two-electron transfer, {\em e.g.}, $\epsilon_l^{(+2)}=u_n(n_x)+u_{n+1}(n_x)-2eV_l$.
The energy transfer rate for two-electron tunneling is then obtained as
\begin{equation}
q_i^{\, n\to n \pm 2}(\epsilon^{(\pm 2)}_i)\ = \ \epsilon^{(\pm 2)}_i
                                       \Gamma_i^{\, n\to n \pm 2}(\epsilon^{(\pm 2)}_i)
\end{equation}
where the reference point is the Fermi level of the leads.

The rates for single-electron transitions are given by a sum of two contributions.
On the one hand, we have the standard expressions for tunneling, {\em e.g.}, 
from a normal to a superconductor~\cite{Abrikosov88} 
\begin{equation}
     \Gamma_i(\epsilon)\ =\ \frac{G_i}{e^2} 2 \int_{\Delta}^{\infty} \mathrm{d}E
                                  \frac{E}{\sqrt{E^2-\Delta^2}}\left[
             f_i(E-\epsilon)f_{\mathrm{isl}}(-E)+f_i(-E-\epsilon)f_{\mathrm{isl}}(E)
                                                               \right]\  , \ i =l,\ r
\end{equation}
(where $f_i(x)=1/(1+\exp{(x/T_i)})$ denotes the Fermi function with the appropriate
temperature $T_l$, $T_r$, or $T$ for the leads or the island, respectively).
On the other hand, there is the {\em escape} rate of a single unpaired 
quasiparticle whose energy equals that of $\Delta$~\cite{Schoen1994,GSZ1994}
\begin{equation}
      \gamma_i(\epsilon)\ =\ \frac{G_i}{e^2}\frac{1}{2\nu_{\mathrm{isl}}}(1-f_i(\Delta+\epsilon))
\end{equation}
with the normal-electron density of states per spin on the island $\nu_{\mathrm{isl}}$.
Note that there is also a corresponding recombination rate. In many cases these
escape rates can be neglected, however, at very low temperatures they may exceed the subgap 
tunneling rate originating from thermally excited quasiparticles in the superconductor.
In that case, they produces a different macroscopic behavior of the system, depending
on whether the total charge number on the island is even or odd.
Thus we have
\begin{equation}
\Gamma_i^{\, n\to n \pm 1}(\epsilon^{(\pm 1)}_i)\ = \ \left\{ \begin{array}{ll}
                          \Gamma_i(\epsilon^{(\pm 1)}_i)& \ \ \ \ \ \mathrm{for}\ n\ \mathrm{even}\\
                          \Gamma_i(\epsilon^{(\pm 1)}_i)+\gamma(\epsilon^{(\pm 1)}_i)
                                                     &   \ \ \ \ \  \mathrm{for}\ n\ \mathrm{odd}\\
                                                              \end{array} \right.
\label{eq:raten_1e}
\end{equation}
where, {\em e.g.}, $\epsilon^{(+1)}_l=u_n(n_x)-eV_l$.
Finally, the corresponding energy transfer rates for single-electron tunneling
are found from 
\[
     \Gamma_i^q(\epsilon)\ =\ \frac{G_i}{e^2} 2 \int_{\Delta}^{\infty} \mathrm{d}E
                                  \frac{E}{\sqrt{E^2-\Delta^2}}\left[
             (E-\epsilon)f_i(E-\epsilon)f_{\mathrm{isl}}(-E)+
            (-E-\epsilon)f_i(-E-\epsilon)f_{\mathrm{isl}}(E)
                                                               \right]
\]
and 
\[
      \gamma_i^q(\epsilon)\ =\ \frac{G_i}{e^2}\frac{\Delta+\epsilon}
                                                   {2\nu_{\mathrm{isl}}}(1-f_i(\Delta+\epsilon))
\]
(where again the reference point is the lead Fermi level) such that we have
\begin{equation}
q_i^{\, n\to n \pm 1}(\epsilon^{(\pm 1)}_i)\ = \ \left\{ \begin{array}{ll}
                          \Gamma^q_i(\epsilon^{(\pm 1)}_i)& \ \ \ \ \ \mathrm{for}\ n\ \mathrm{even}\\
                          \Gamma^q_i(\epsilon^{(\pm 1)}_i)+\gamma^q(\epsilon^{(\pm 1)}_i)
                                                     &   \ \ \ \ \  \mathrm{for}\ n\ \mathrm{odd}\\
                                                              \end{array} \right.
\label{eq:Qraten_1e}
\end{equation}
In order to calculate the currents~\eqref{eq:ecurrent}, \eqref{eq:qcurrent} 
we solve the stationary master equation for the probabilities
$P_n$
\[
     \frac{\partial P_n}{\partial t}\ = \ 0\ =\ 
                                    \sum_{k\neq n} P_k \Gamma^{\, k \to n}-P_n \Gamma^{\, n \to k} 
\]
for an applied (small) bias voltage $\Delta V$ or a (small) temperature difference $\Delta T$.
If a voltage or temperature difference is applied the island is, strictly speaking, in a
non-equilibrium state. For the given physical situation it is reasonable to neglect
the non-equilibrium part of the distribution function (which we have already done
by writing the transfer rates in the form above). For the temperature $T$ of the island we 
assume the arithmetic mean $T=(T_l+T_r)/2$. We note that there is an independent
test of the numerical calculation by checking the Onsager relation $G_T=M/T$ for the
coefficients in Eq.~\eqref{eq:lintransport} which is obeyed to a high accuracy 
by our method. 

\section{The parity effect}
As mentioned above, the parity effect arises as a macroscopic manifestation
of the parity of the electron number in a superconductor, {\em i.e.}, 
different behavior depending on whether the total electron number is even or odd.
The effect was predicted in Ref.~\cite{AverinNazarov1992} and first 
observed by Tinkham {\em et.\ al.}~\cite{Tinkham1992}.

In an even-number superconductor at $T=0$ all electrons near the Fermi level
are bound in Cooper pairs and there is not a single unpaired quasiparticle
left. This system has an energy gap $2\Delta$. On the other hand, adding
{\em one} electron results in a single quasiparticle excitation which
which does not have an excitation gap. However, the response to an external
field is drastically reduced as it is caused just by a single electron.
It is intuitively clear that the difference between the two parities 
will fade away as soon as there are more thermal quasiparticles.
This defines the criterion for a crossover temperature $T^{\ast}$ beyond
which even-odd differences disappear for an isolated superconductor such
as the island in the NSN SET: It is the temperature at which there
is on average one thermally excite quasiparticle and
it is defined  by
\begin{equation}
\label{eq:crossoverT}
T^*\ =\ \frac{\Delta}{\ln{N_{\mathrm{eff}}(T^*)}}
\end{equation}
where
$N_{\mathrm{eff}}(T)=\nu_{\mathrm{isl}}\sqrt{2\pi T\Delta}$ can be viewed as 
an effective number of accessible quasiparticle states  at the temperature
$T$. Superconducting islands in SET are made of aluminum, and for typical
 parameters (cf.~Refs.~\cite{Tinkham1992,Eiles1993,Hergenrother1994}) one has
$N_{\mathrm{eff}}\sim 10^4$ and $T^*\sim 250$mK.

Keeping in mind that an SET has the characteristic energy scale $E_C$,
there arise four interesting transport regimes: For the gap  and
the charging energy we may have $\Delta < E_C$ or $\Delta > E_C$. These
two cases may be studied for $T>T^*$ (no parity effects) or in the
parity regime $T< T^*$. Interestingly, with our numerical method 
we have a choice for studying the system with parity effects.
For computations at high temperatures $T>T^*$ it does not matter whether
or not the two-electron and escape rates are included, they
do not give any observable effect.
On the other hand, for $T<T^*$  even-odd differences cannot be observed 
without including these rates. Therefore, including or not including
these rates in calculations below $T^*$ helps us to identify the 
contribution due to the 'parity-generating' processes.
%
%

\section{Thermopower of an NSN setup}

Before we discuss our results we briefly recall an idea
due to Matveev~\cite{Matveev00} (cf.\ also~\cite{Turek2005})
to interpret the thermopower as an average energy $\langle \xi \rangle$
at which
current is transported in the voltage-biased system (under
linear transport conditions). This is an intuitive and 
powerful method which will use throughout our work
to interpret our results.

The idea can be understood by taking into account that the current $I_e$
through a device can be written as 
$I_e=-e\int [f_l(\xi)-f_r(\xi)]w(\xi){\mathrm d}\xi$
where $f_l(\xi)$ and $f_r(\xi)$ denote the Fermi functions in
the left and right lead for the respective potential and temperature
(the energies $\xi$ are taken with respect to the chemical potentials).
Here $w(\xi)$ includes all other quantities such as density
of states in the leads and transparency of the device at the energy $\xi$.
By noting that $G_T=\partial I_q/\partial T_l$ and 
               $G_V=\partial I_e/\partial V$ one finds
\[
   \langle \xi \rangle \equiv
            \frac{\int\xi\left(-\frac{\partial f}{\partial\xi}\right)w(\xi){\mathrm{d}}\xi}
                 {\int   \left(-\frac{\partial f}{\partial\xi}\right)w(\xi){\mathrm{d}}\xi}
    = \frac{(-e)T G_T}{G_V}
\]
and consequently for the thermopower
\begin{equation}
\label{eq:average_energy}
S \ = \ - \frac{\langle \xi \rangle}{e T} \ \ .
\end{equation}
That is, up to a factor the thermopower measures directly this average energy 
$\langle \xi \rangle$.

Let us turn now to our calculations for the thermopower of NSN SET.
For the numerics we measure energy and inverse time in units of $E_C$,
and conductance in units of $e^2$. Then, the free parameters are
the ratios of the conductances $G_l$, $G_r$, $G_{A,l}$, $G_{A,r}$,
and the escape rates $\gamma_l$, $\gamma_r$.
The realistic parameter values which enter our calculations are: 
$G_l=G_r=(50\mathrm{k}\Omega )^{-1}$,
$G_{A,l}=G_{A,r}=5\cdot 10^{-9}\Omega^{-1}$, $E_C=100\mu$eV, 
$\gamma_l=\gamma_r=10^6 \mathrm{s}^{-1}$. The experiments in
Ref.~\cite{Eiles1993,Hergenrother1994} have been carried out
with device parameters close to these values.

\subsection{$\Delta < E_C$}
In Fig.~2 we show the results for calculations of
the thermopower with and without parity effects.
The functional dependence $S(n_x)$ in the latter case
has been explained analytically in Ref.~\cite{Turek2005}.
As we have $T\ll \Delta, E_C$ it is sufficient to 
include two (or at most three) charge states in the 
considerations in  order
to understand the behavior of $S(n_x)$.
Here we focus on a brief discussion of the additional 
features in the parity regime $T<T^*$.
\begin{figure}[h]
\vspace*{3mm}
\centerline{
\epsfxsize=0.43\textwidth
\epsfbox{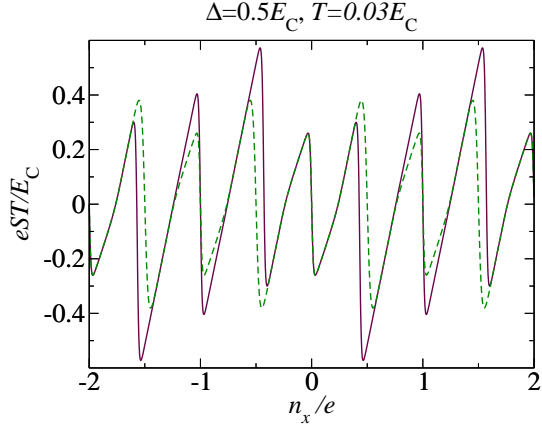}}
\label{fig2}
\caption{
 Thermopower $S(n_x)$ of the NSN SET with $\Delta = 0.5 E_c$
 and temperature $T=0.03 E_C$. The green dashed line
 represents the thermopower without two-electron tunneling
 and escape rate and therefore displays no even-odd differences.
 The purple solid line displays the result including all
 processes. It shows clear $2e$ periodicity and an enhancement
 of the thermopower for certain gate charges.
 }
\end{figure}
The key to understand the functional dependence of $S(n_x)$
without even-odd effects is that at the points $u_n=0$,
that is, when $n_x$ takes half-integer values tunneling
to the island $n\to n+1$ and from the island $n+1 \to n$
occurs with equal probability. By using simple arguments
for the probabilities $P_n$ and $P_{n+1}$ the following
equation has been derived in Ref.~\cite{Turek2005}
\begin{equation}
\label{eq:estimate1_S}
 S \ =\  - \frac{1}{e T} \left( u_0(n_x) - 
   \Delta \tanh \left[ \frac{u_0(n_x)}{2 T} \right] \right) 
\end{equation}
which governs the behavior of $S(n_x)$ in the intervals from
$u_n(n_x)=0$ to the zeros of $S(n_x)$ which are closest to
those $n_x$ values.

By inspecting Fig.~2 we note that the essential difference
introduced in the parity regime is that the
purple curve is shifted from the half-integer values of
$n_x$ towards the closest even number. As the slope
of the curve does not change in this shift, the consequence
is that the maximum absolute value of $S$ increases,
compared to the case without parity.
It is not difficult to describe this behavior analytically
by repeating the arguments that lead to Eq.~\eqref{eq:estimate1_S},
now taking into account that the recombination rate of
quasiparticles on the island is no longer $\propto \exp{(-\Delta/T)}$.
In this range of $n_x$ there is an unpaired quasiparticle on the
island whose recombination rate provides the dominating contribution
to the charge current. 
This rate is 
$ \propto  (e^{-\Delta/T} +  \frac{1}{2N_{\mathrm{eff}}}) $.
From this we get a modified relation
\begin{equation}
\label{eq:modestimate_S}
 S \ =\  - \frac{1}{e T} \left( u_0(n_x) - 
   \Delta \tanh \left[ \frac{u_0(n_x)+\eta}{2 T} \right] \right) 
\end{equation}
with $\eta= T\ln{\left[1+\exp{\left(\frac{\Delta}{T}+\ln{\frac{1}{2N_{\mathrm{eff}}}
                                                        }
                              \right)}
                 \right]
                }$. The relation~\eqref{eq:modestimate_S}
captures the essential features of the purple curve in Fig.~2.

\subsection{$\Delta > E_C$}

The green dashed line in Fig.~3 shows an example for the thermopower 
without parity effects for $\Delta > E_C$. The arguments we have
given in the preceding subsection for the thermopower at half-integer
values of $n_x$ are valid also in this case. That is, 
Eq.~\eqref{eq:estimate1_S} correctly describes the behavior of the function
$S(n_x)$ also in this parameter range. Hence, this equation turns out
to be the key for understanding the thermopower in NSN SET, {\em i.e.}, 
a device with an electronic spectrum that is gapped around the Fermi energy.
We further mention that for the thermopower without parity effects
for the average energy close to integer values of $n_x\simeq n$ one has
$\langle \xi \rangle \simeq (u_n+u_{n-1})/2$ which yields
\begin{equation}
          S\ \simeq\ -\frac{u_n+u_{n-1}}{2eT} \ \ ,\ \ n_x\simeq n\ \ .
\label{eq:Delta_groesser_Ec}
\end{equation}
\begin{figure}[hb]
\vspace*{3mm}
\centerline{
\epsfxsize=0.43\textwidth
\epsfbox{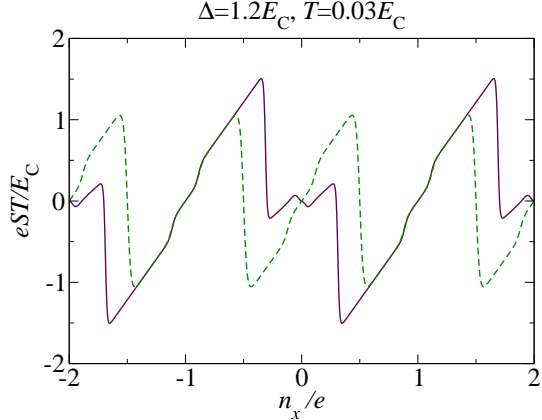}}
\label{fig3}
\caption{
 Thermopower $S(n_x)$ of the NSN SET with $\Delta = 1.2 E_c$ 
 and $T=0.03 E_C$ a) including all tunneling processes (purple
 solid line), b) without two-electron tunneling and without
 escape processes (green dashed line). While the green line
 is $e$-periodic the purple curve shows clear $2e$-periodicity
 and substantial qualitative changes with respect to the
 case without even-odd effect.
 }
\end{figure}

If the gap exceeds the charging energy, two-electron tunneling
starts to play a prominent role for the current-voltage characteristics
of NSN SET in the parity regime $T<T^*$~\cite{Hekking1993}. 
Around odd-integer values of $n_x$ there occurs a current peak 
which is due to a cycle of two-electron tunneling processes.
Note that these processes do not have a gap. The thermopower
is analogous to that of single-electron tunneling in NNN SET,
the only difference is that we have to substitute the
charging energy difference for two-electron tunneling, 
$\langle \xi \rangle \simeq (u_n(n_x)+u_{n-1}(n_x))/2$ which leads
to a thermopower $S\simeq 
          -(u_n+u_{n-1}/(2eT) $. This result curiously coincides
with the one in the absence of parity effects, 
Eq.~\eqref{eq:Delta_groesser_Ec} although the dominating 
tunneling process is a different one.

In the vicinity of the half-integer values of $n_x$ we observe
the analogous effect to the case $\Delta < E_C$: The curve
gets shifted towards the closest even-integer $n_x$ without
changing the slope. Scrutiny of the dominating current-carrying
processes, {\em e.g.}, for $n_x < 1/2$ reveals that also here 
the largest rate is due to quasiparticle recombination on 
the island. However, if $T<T^*$ there are essentially
no thermal quasiparticles. There is only a {\em single}
unpaired electron which is left from the pair-breaking
tunneling off the island. The quantitative description of
this recombination process leads in full analogy to the
conclusion of the previous subsection, namely that  
Eq.~\eqref{eq:modestimate_S} describes this shift of the 
curve. The essence also here is that the current is due to
tunneling of a single unpaired quasiparticle.
%

\section{Figure of merit {\em ZT}}
Once we have studied the thermopower of the NSN SET
it is an interesting question to investigate the
thermoelectric efficiency of this device.
This efficiency  is quantified by 
\begin{equation}
    ZT\ =\ \frac{G_V S^2 T}{\kappa}\ =\ \frac{G_V S^2 T}{\kappa_e+\kappa_l}
\label{eq:ZT}
\end{equation}
where $\kappa_e$ and $\kappa_l$ denote the electronic and lattice contribution to the heat
conductance of the device. Our work focuses on a regime of extremely
low temperatures for which $\kappa_l$ is very small, hence we neglect it and 
$\kappa=\kappa_e$.
We compute $\kappa$ from the heat current $I_q$ 
according to Eqs.~\eqref{eq:lintransport}--\eqref{eq:qcurrent}.

An example of the results is shown in Fig.~4. 
For $\Delta \sim E_C$ we observe huge values of $ZT$ 
which is rather uncommon, keeping in mind that typical
\begin{figure}[h]
\vspace*{3mm}
\centerline{
\epsfxsize=0.43\textwidth
\epsfbox{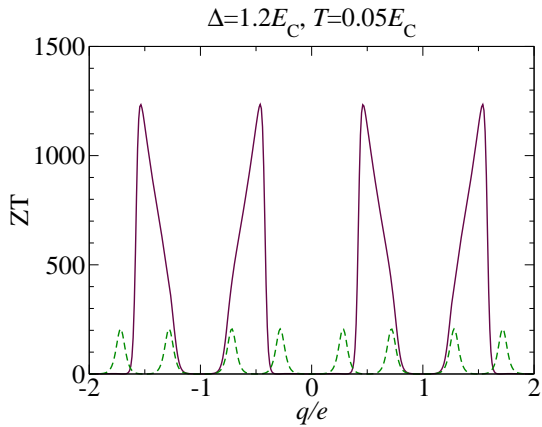}}
\label{fig4}
\caption{
 Figure of merit $ZT$ for an NSN SET with $\Delta = 1.2 E_c$ 
 and $T=0.03 E_C$ a) including all tunneling processes (purple
 solid line), b) without two-electron tunneling and without
 escape processes (green dashed line). A strong enhancement
 of $ZT$ is found in a range of $n_x$ values  where also the 
 modulus of the thermopower
 reaches its maximum (cf.~Fig.~3).
 }
\end{figure}
values for materials reach the order of 1, or for quantum dots the
order of $10^2$. The question is why such high values are possible for
this system. A quick answer might be that, given that superconductors
are bad heat conductors, it could be expected that by including
the superconducting island $ZT$ of a single-electron transistor 
would be enhanced. Let us try to give a more quantitative answer
which takes into account our observations from the previous section.

By transferring Matveev's idea for the interpretation of $S$ as an average
transport energy of the electrons according to
Eq.~\eqref{eq:average_energy} we can derive 
an expression that illuminates the meaning of $ZT$. In analogy with 
Eq.~\eqref{eq:average_energy} we obtain $K=\langle \xi^2 \rangle G_V/(e^2 T)$.
This leads, together with Eq.~\eqref{eq:ZT} and
$\kappa=K-G_VTS^2$, to the new relation
\begin{equation}
         ZT\ = \ \frac{\langle \xi \rangle^2}{\langle \xi^2 \rangle-\langle \xi \rangle^2}
      \ \ .
\label{eq:ZT_new}
\end{equation}
Indeed, all additional factors cancel  and $ZT$ turns out to be the ratio
of the squared average transport energy and the variance of that energy.
This relation clearly indicates the strategy that needs
to be used in order to increase $ZT$: The current-carrying electrons should
be far from the Fermi energy while their energetic distribution should be
as narrow as possible. This corroborates also the conclusion of Ref.~\cite{Mahan1996}
that a $\delta$ function in the spectrum is favorable for a high $ZT$ value.

Let us now apply Eq.~\eqref{eq:ZT_new} to the NSN SET. We have already discussed
that for the $n_x$ values where the strong enhancement of $ZT$ is found, 
the current is carried by a single unpaired quasiparticle. The maximum
energy of this quasiparticle is $\sim \Delta$ while its energy distribution is rather 
narrow: it is just given by the temperature $T$. Hence we expect 
$ZT\sim \Delta^2/T^2$. In fact, this estimate has the correct order of magnitude.

\section{Discussion}

We have investigated the thermopower for a single-electron transistor
with a superconducting island for arbitrary ratios $\Delta/E_C$ and
for temperatures both above and below the crossover temperature for
parity effects $T^*$. The results show the expected parity effects
also in the functional dependence $S(n_x)$ of the thermopower on the gate
charge $n_x$, in particular $2e$ periodicity. We have provided a
discussion of the essential features of this functional dependence
in terms of Eqs.~\eqref{eq:modestimate_S}, \eqref{eq:Delta_groesser_Ec}.
It is remarkable that the basis for this discussion is Eq.~\eqref{eq:estimate1_S}
which was found already in Ref.~\cite{Turek2005}.

Apart from the thermopower we have also calculated the thermoelectric figure
of merit $ZT$. Unexpectedly we have found a strong enhancement of $ZT$
compared to common values of this quantity. In order to understand our
findings we have given a new interpretation of $ZT$ in terms of Matveev's
idea to represent thermoelectric quantities as moments of the 
energy distribution for the current-carrying electrons. It shows that
large values of $ZT$ can be obtained if the dominant transport mechanism
occurs far from the Fermi level, and at the same time, has a narrow
distribution in its energies. Clearly, the NSN SET is a system where
these conditions can be achieved.

\section{Acknowledgments} 
Financial support through Priority program 1386
of the German Research Foundation
is gratefully acknowledged. 


\end{document}